\documentclass{iaus}
\usepackage{graphicx}

\title[He-accreting WDs as SNe Ia Progenitors] 
{He-accreting WDs as SNe Ia Progenitors}

\author[Piersanti, Tornamb\'e, Yungelson \& Straniero]   
{Luciano Piersanti$^1$,
 Amedeo Tornamb\'e$^2$
 Lev Yungelson$^3$
 \and Oscar Straniero$^1$}

\affiliation{$^1$INAF - Osservatorio Astronomico di Teramo, \\ 
via Mentore Maggini, snc, 64100, Teramo, ITALY \\
email: {\tt piersanti,straniero@oa-teramo.inaf.it} \\[\affilskip]
$^2$INAF - Osservatorio Astronomico di Roma, \\ 
via di Frascati, 33, 00040, Monteporzio Catone, ITALY \\
email: {\tt tornambe@oa-teramo.inaf.it} \\[\affilskip]
$^3$Institute of Astronomy, Moscow, RUSSIA \\
email: {\tt lev.yungelson@gmail.com}}

\pubyear{2011}
\volume{281}  
\pagerange{119--122}
\jname{Binary Paths to Type Ia Supernovae Explosions}
\editors{A.C. Editor, B.D. Editor \& C.E. Editor, eds.}
\begin{document}

\maketitle

\begin{abstract}
We investigate the thermal response of CO WDs accreting He-rich matter directly from their companions 
in binary systems. Our results suggest that the He-channel cannot provide progenitors for majority of  
``normal'' SNe Ia. 
\end{abstract}

\firstsection 
\section{Introduction}

Observational evidence clearly suggests that type Ia Supernovae (SNe Ia) are 
produced by the thermonuclear disruption of CO WDs accreting matter from their 
companions in binary systems. However, up to now no clear consensus does exist concerning 
the nature of the donor. Currently the most favored scenarios involve
either a normal star with a H-rich envelope (Single Degenerate scenario - \cite{whelan73})
or another CO WD (Double Degenerate scenario - \cite{ibentut84,web84}). 
An alternative evolutionary path is the so called He-donor channel; in this case He-rich matter 
is directly accreted from a He-star or from a He-WD. These objects are very interesting as, 
in principle, they could produce an explosion of SN Ia scale both via C-deflagration 
in a CO WD close to the Chandrasekhar limit and He-detonation in a Sub-Chandrasekhar mass WD. 
Even if during the last decades He-detonation has not been considered promising as SNe Ia 
explosion mechanism ({\it e.g.} see \cite{ww94}), recently \cite{sim2010} have demonstrated 
that, if the CO core is larger than $\sim 0.9 M_\odot$ and the He-buffer at the onset of the 
dynamical flash is small enough ($\leq 0.1 M_\odot$), the resulting explosion could resemble 
all the observational properties of a ``normal'' SN Ia. 
\cite{ruiter2011} have investigated by means of population synthesis technique 
the expected frequency of He-accreting WD evolving into a SN Ia-like event and they 
concluded that the Sub-Chandrasekhar ``{\sl ... model is the first model which demonstrates a 
sufficient number of SNe Ia events to account for all, or at least some substantial fraction of, 
SNe Ia ..., as well as two distinct formation channels with their own characteristic DTD.}''

Even if a large effort has been devoted in the past to investigate the 
thermal response of CO WDs accreting directly He-rich matter (a list of works dealing 
with this topic can be found in \cite{shen2009}), an overall fully evolutionary scenario is still 
missing. For this reason we computed a large set of fully evolutionary models of 
CO WDs in the mass range $0.6\div 1.1 M_\odot$ accreting He-rich matter with $\dot{M}$ in the 
range $10^{-9}\div 10^{-5} M_\odot yr^{-1}$. 
The aim of our work is twofold: 1) the definition of the possible accretion regime, and 
2) the final outcome of the accretion process.

\begin{figure}[t]
\begin{center}
 \includegraphics[width=4in]{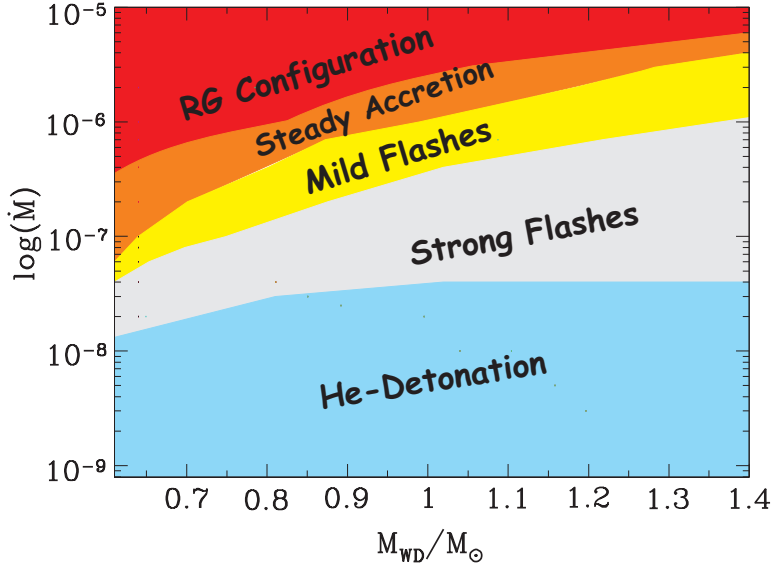} 
 \caption{Possible accretion regimes in the $M_{WD} - \dot{M}$ plane.}
   \label{piersanti_fig1}
\end{center}
\end{figure}
\section{Results} 
Our results are summarized in Fig. \ref{piersanti_fig1} where we report as a function of 
the WD total mass and accretion rate the possible accretion regimes.

\begin{figure}[t]
\begin{center}
 \includegraphics[width=3.4in]{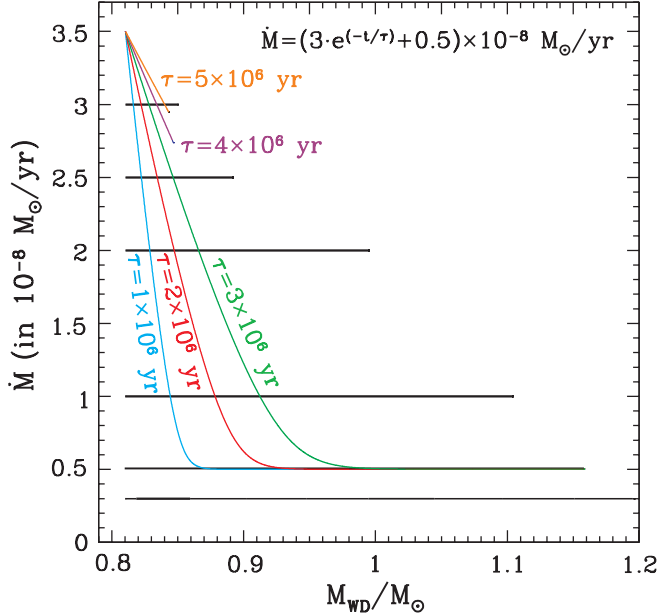} 
 \caption{Mass extension of the He-buffer for a 0.8 $M_\odot$ CO WD accreting at various accretion rates. 
          For more details see text.}
   \label{piersanti_fig2}
\end{center}
\end{figure}

\begin{itemize}
\item {\bf RG Configuration}: for very high value of the $\dot{M}$, but below the Eddington limit, 
the accretion rate is larger than the rate at which matter is converted into CO. Hence, a massive He-buffer 
forms and the WD expands to giant dimension, overfilling its own Roche lobe. This implies that part of the 
transferred mass is ejected from the WD and it is lost from the binary system. The further evolution of 
systems strongly depends on the amount of angular momentum carried away by the lost matter. If the ejected 
matter has the same specific angular momentum as the accretor, the mass transfer is dynamically stable 
but non conservative, while for larger value it becomes dynamically unstable and it evolves in a Common Envelope. 
In the former case part of the transferred mass is effectively accreted onto the WD, while in the 
latter the accumulation efficiency is practically zero ($\eta\simeq 0$). 
This accretion regime occurs at the beginning of the mass transfer process when the 
donor is a He-WD or a massive He-star ($M_{He-star}> 0.85 M_\odot$). 

\item{\bf Steady Accretion}: $\dot{M}$ is equal to the rate at which He is converted into CO.
All the transferred mass is effectively accreted onto the WD which can grow in mass up to the 
Chandrasekhar limit.

\item{\bf Mild Flashes}: He-burning is almost switched off, a 
He-buffer is piled up and when it exceeds some critical value (depending both on $M_{CO}$ and $\dot{M}$) 
a He-flash is ignited. For $\dot{M}$ typical of this accretion regime the flash is mild and, hence, 
the expansion triggered by the sudden release of energy is small so that the WD 
remains well inside its own Roche lobe. As a consequence the transferred matter is fully 
accumulated onto the WD.

\item{\bf Strong Flashes}: as in the previous accretion regime, the mass transfer 
proceeds through recurrent He-flashes, but in this case the energy delivered is so large that 
the accreting WD expands to giant dimension. As a consequence a part of the previously accreted 
matter is ejected from the WD and lost by the system. 
The accumulation efficiency for this kind of systems is very low, but it could largely increase 
if the effects of strong radiation-driven wind are taken into account (\cite{kato2004}). For example, 
for a model with $M_{WD}=1.02 M_\odot$ and $\dot{M}=4\times 10^{-7} M_\odot yr^{-1}$ we obtain 
$\eta=0.11$ while \cite{kato2004} provide $\eta=0.77$. 

\item{\bf He-Detonation}: for $\dot{M}\le 2\times 10^{-8} M_\odot yr^{-1}$, almost independently of 
the the initial WD mass, the physical base of the 
He-rich zone strongly degenerates and, when the accreted buffer exceeds a critical value, 
a dynamical He-flash is ignited. The resulting He-detonation triggers the explosion of the accreting WD. 
The mass extension of the He-buffer is almost independent of the WD mass but it strongly depends on the 
accretion rate: lower the accretion rate larger the mass extension of the He-buffer at the onset of 
the dynamical flash. This is shown in Fig. \ref{piersanti_fig2} where we report for a 0.8 $M_\odot$ 
CO WD the final mass at the explosion as a function of 
$\dot{M}$. Moreover, $\Delta M_{He}$ strongly depends on the time evolution of $\dot{M}$, as shown 
in the same plot where we report the evolution of CO WD models accreting He-rich matter 
with a time-dependent law.
\end{itemize}
\begin{figure}[t]
\begin{center}
 \includegraphics[width=3.4in]{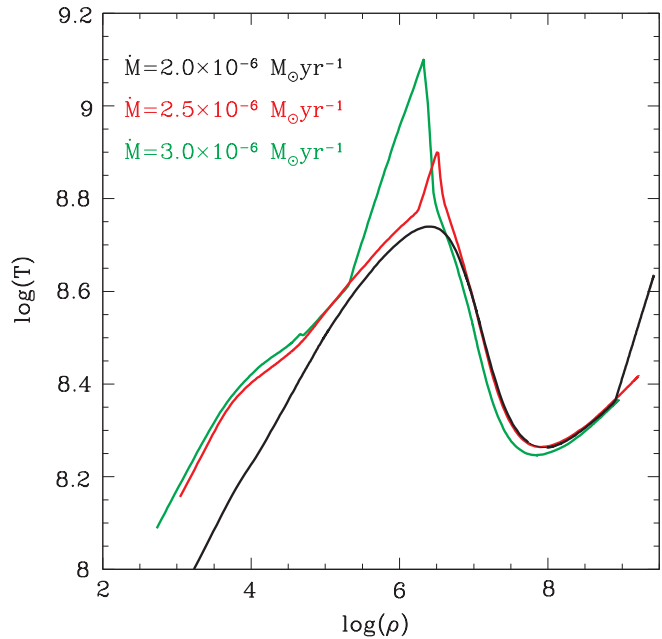} 
 \caption{Profile in the $\rho - T$ plane at the C-ignition for models with different 
$\dot{M}$.}
   \label{piersanti_fig3}
\end{center}
\end{figure}

As we follow the long-term evolution of He-accreting WDs, we can also determine whether or not 
a given system will attain the Chandrasekhar limit, experiencing a C-deflagration. In Fig. \ref{piersanti_fig3} we report
the profiles in the $\rho - T$ plane at the C-ignition for models accreting He-rich matter onto an 
initial $1.02 M_\odot$ CO WD at various rates, as labeled. Even if all the models have almost the 
same final mass ($M_{WD}\sim 1.36 M_\odot$), for $\dot{M}=2\times 10^{-6} M_\odot yr^{-1}$ central C-ignition occurs, while for higher values 
C-burning is ignited off-center, very close to the stellar surface. In the former case a SN Ia event will result, while in the latter C-burning 
will propagate inward steadily, thus producing an ONeMg WD, which could eventually experience core collapse, thus forming a 
neutron star (\cite{saio2004}).


Our results suggest that the expected frequency of He-detonating systems with a large CO core and a small He-buffer is much lower than the 
one estimated by Ruiter and coworkers. In fact, the accumulation efficiency in the RG Configuration regime is smaller than 1,thus limiting 
the growth in mass of the CO core. Moreover, the mass extention of the He-buffer at the onset of the dynamical He-flash does depend on 
the accretion history. At the end, the area in the parameter space $M_{WD}-\dot{M}$ for which a CO WD could ignite C-burning at the center 
when it attains the Chandrasekhar limit is smaller than previously suggested ({\it e.g.} \cite{nomoto1982}). In fact, for accretion rate 
larger than $2\times 10^{-6} M_\odot yr^{-1}$ C-burning is ignited off-center thus producing a collapsing ONeMg, not a SN Ia event.

\begin{discussion}

\discuss{Sim}{According to your results what is the expected frequency of He-detonating WDs 
with large CO core and small He-buffer?}

\discuss{Piersanti}{I can not provide the exact value as we do not use our results as input for a population 
synthesis simulation, but I am quite confident that the frequency for ``good'' He-detonating WDs 
is at least a factor 50 (or more) smaller than the one estimated by Ruiter and coworkers.}

\end{discussion}

\end{document}